\documentclass{article}
\usepackage{dcase2022,amsmath,amsfonts,graphicx,url,times,booktabs,tabularx,multirow}
\usepackage[table,xcdraw]{xcolor}

\newcommand{\norm}[1]{\left\lVert#1\right\rVert}
\usepackage{soul}
\usepackage{floatrow}

\usepackage{hyperref}
\newfloatcommand{capbtabbox}{table}[][\FBwidth]

\usepackage{blindtext}

\title{Improving Natural-Language-based Audio Retrieval \\ with Transfer Learning and Audio \& Text Augmentations}
\name{Paul Primus$^1$, Gerhard Widmer$^{1,2}$}
\address{
$^1$Institute of Computational Perception (CP-JKU)  \\
$^2$LIT Artificial Intelligence Lab\\
Johannes Kepler University, Austria
}

\begin{document}

\ninept
\maketitle

\begin{sloppy}

\begin{abstract}
The absence of large labeled datasets remains a significant challenge in many application areas of deep learning. Researchers and practitioners typically resort to transfer learning and data augmentation to alleviate this issue. We study these strategies in the context of audio retrieval with natural language queries (Task 6b of the DCASE 2022 Challenge). Our proposed system uses pretrained embedding models to project recordings and textual descriptions into a shared audio-caption space in which related examples from different modalities are close. We employ various data augmentation techniques on audio and text inputs and systematically tune their corresponding hyperparameters with sequential model-based optimization. Our results show that the used augmentations strategies reduce overfitting and improve retrieval performance.
\end{abstract}

\begin{keywords}
Language-based Audio Retrieval, Transfer Learning, Audio Augmentation, Text Augmentation
\end{keywords}

\section{Introduction}
\label{sec:intro}
Natural-language-based audio retrieval is concerned with ranking audio recordings depending on their content's similarity to textual descriptions. Retrieval tasks like this are typically solved by converting recordings and textual descriptions into high-level representations and then aligning them in a shared audio-caption space; ranking can then be done based on the distance between embeddings. These systems' retrieval performance highly depends on the quality of the audio and text embedding models, which must extract features that accurately and discriminatively represent the high-level content. Current state-of-the-art approaches \cite{recent_work_1, recent_work_2, clap} create such feature extractors by training models with millions of parameters directly from raw input features, i.e., deep learning. These large embedding models require a large number of training examples, such as the 400 million image-text pairs used to train CLIP \cite{clip}, a cutting-edge image-retrieval model. However, publicly available audio-caption datasets like Clotho and AudioCaps are significantly smaller. This work showcases how to use off-the-shelf pretrained audio and text neural networks to create a state-of-the-art retrieval model under this limiting condition. We evaluate our approach in the context of task 6b\footnote{ \url{https://dcase.community/challenge2022/task-language-based-audio-retrieval}} of the 2022's DCASE Challenge  \cite{dcase2022_task6b}, which is concerned with audio retrieval from natural language descriptions. We demonstrate how an already well-performing baseline model can be further improved by using a range of audio and text augmentation methods and pretraining on AudioSet.
\begin{figure}[t]
    \centering
    \def\svgwidth{0.70 \columnwidth}
\begingroup%
  \makeatletter%
  \providecommand\color[2][]{%
    \errmessage{(Inkscape) Color is used for the text in Inkscape, but the package 'color.sty' is not loaded}%
    \renewcommand\color[2][]{}%
  }%
  \providecommand\transparent[1]{%
    \errmessage{(Inkscape) Transparency is used (non-zero) for the text in Inkscape, but the package 'transparent.sty' is not loaded}%
    \renewcommand\transparent[1]{}%
  }%
  \providecommand\rotatebox[2]{#2}%
  \newcommand*\fsize{\dimexpr\f@size pt\relax}%
  \newcommand*\lineheight[1]{\fontsize{\fsize}{#1\fsize}\selectfont}%
  \ifx\svgwidth\undefined%
    \setlength{\unitlength}{282.62786848bp}%
    \ifx\svgscale\undefined%
      \relax%
    \else%
      \setlength{\unitlength}{\unitlength * \real{\svgscale}}%
    \fi%
  \else%
    \setlength{\unitlength}{\svgwidth}%
  \fi%
  \global\let\svgwidth\undefined%
  \global\let\svgscale\undefined%
  \makeatother%
  \begin{picture}(1,1.0280965)%
    \lineheight{1}%
    \setlength\tabcolsep{0pt}%
    \put(0,0){\includegraphics[width=\unitlength,page=1]{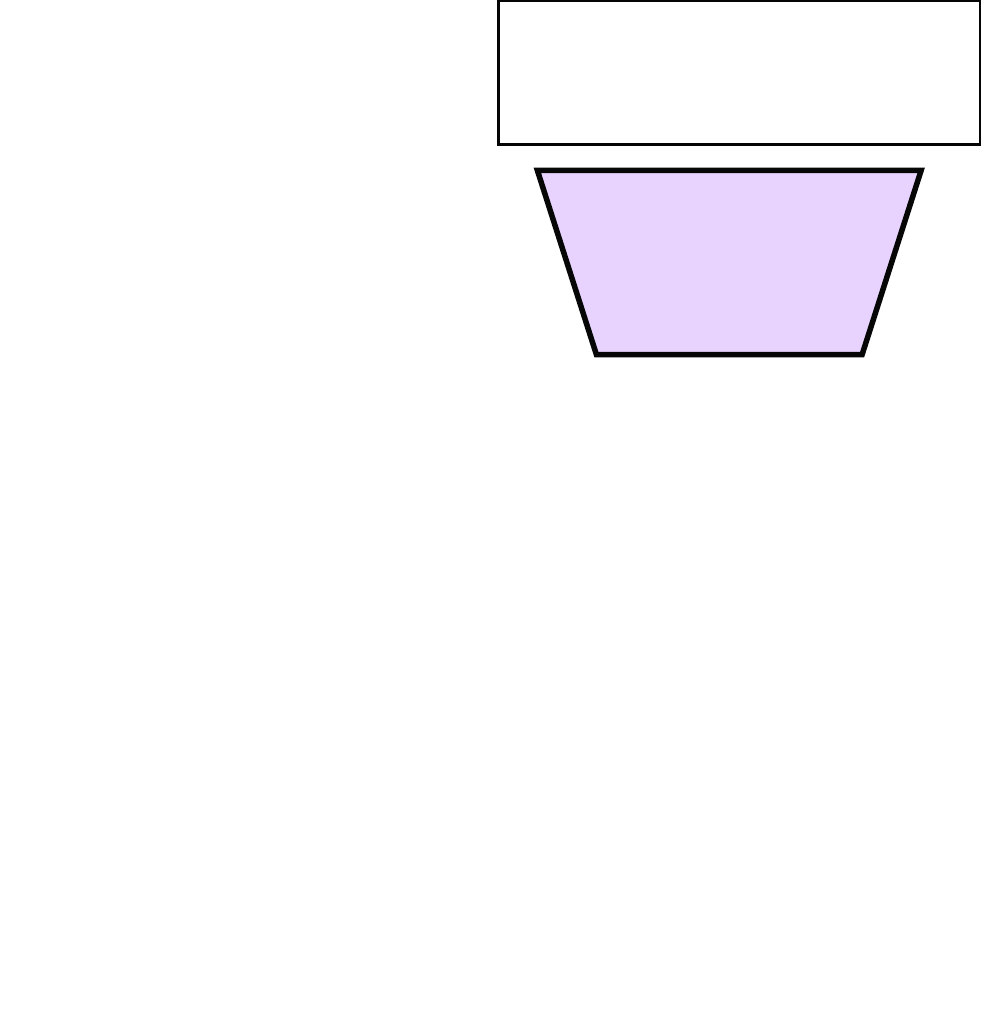}}%
    \put(0.52247715,0.97276377){\color[rgb]{0,0,0}\makebox(0,0)[lt]{\lineheight{1.25}\smash{\begin{tabular}[t]{l}A heavy rain storm is \\coming down outside.\end{tabular}}}}%
    \put(0,0){\includegraphics[width=\unitlength,page=2]{embedding.pdf}}%
    \put(0.21187225,0.76155104){\color[rgb]{0,0,0}\makebox(0,0)[lt]{\lineheight{1.25}\smash{\begin{tabular}[t]{l}$\phi_a$\end{tabular}}}}%
    \put(0.71448746,0.75959948){\color[rgb]{0,0,0}\makebox(0,0)[lt]{\lineheight{1.25}\smash{\begin{tabular}[t]{l}$\phi_c$\end{tabular}}}}%
    \put(0,0){\includegraphics[width=\unitlength,page=3]{embedding.pdf}}%
    \put(0.83524437,0.00834454){\color[rgb]{0,0,0}\makebox(0,0)[lt]{\lineheight{1.25}\smash{\begin{tabular}[t]{l}$d_1$\end{tabular}}}}%
    \put(-0.002937,0.44156939){\color[rgb]{0,0,0}\makebox(0,0)[lt]{\lineheight{1.25}\smash{\begin{tabular}[t]{l}$d_2$\end{tabular}}}}%
    \put(0,0){\includegraphics[width=\unitlength,page=4]{embedding.pdf}}%
  \end{picture}%
\endgroup%

    \caption{The proposed audio-retrieval system in a nutshell: Audio and descriptions are transformed into the shared audio-caption embedding space via the audio and description embedding models $\phi_\mathrm{a}$ and $\phi_\mathrm{c}$, respectively. The contrastive loss maximizes the similarities between matching pairs.}
    \label{fig:overview}
\end{figure}

\section{Related Work}
The idea of aligning text and audio features for content-based retrieval is not new: Early audio retrieval methods connected bag-of-words text queries and MFCC features via density or discriminative models \cite{previous_work_2}. However, the handcrafted features and the relatively small vocabulary limited these methods' performance. Current methods build on top of learnable feature extractors that produce high-level audio and text representations from raw input features. Xie et al. \cite{recent_work_2}, for example, used a convolutional recurrent neural network to extract frame-wise acoustic embeddings and aligned those to Word2Vec features via a linear transformation. Recently, language-based audio retrieval has received increased attention due to the newly introduced task 6b in the 2022's DCASE challenge \cite{dcase2022_task6b}. The task's objective was to create a retrieval system that takes natural-language queries as input and retrieves the ten best-matching recordings from a test set. The top ranking systems among the nine submitted ones leveraged large pretrained audio and text embedding models like CNN14 \cite{panns} and BERT \cite{bert}, respectively. While most systems applied SpecAugment \cite{specaugment}, other data augmentation methods, especially text augmentations, have received little to no attention. We address this paucity and study a range of audio and text augmentation methods in the context of audio retrieval.

\begin{figure*}[t!]

\begin{floatrow}
\ffigbox{%
      
    \def\svgwidth{.9 \columnwidth}
\begingroup%
  \makeatletter%
  \providecommand\color[2][]{%
    \errmessage{(Inkscape) Color is used for the text in Inkscape, but the package 'color.sty' is not loaded}%
    \renewcommand\color[2][]{}%
  }%
  \providecommand\transparent[1]{%
    \errmessage{(Inkscape) Transparency is used (non-zero) for the text in Inkscape, but the package 'transparent.sty' is not loaded}%
    \renewcommand\transparent[1]{}%
  }%
  \providecommand\rotatebox[2]{#2}%
  \newcommand*\fsize{\dimexpr\f@size pt\relax}%
  \newcommand*\lineheight[1]{\fontsize{\fsize}{#1\fsize}\selectfont}%
  \ifx\svgwidth\undefined%
    \setlength{\unitlength}{372.54399099bp}%
    \ifx\svgscale\undefined%
      \relax%
    \else%
      \setlength{\unitlength}{\unitlength * \real{\svgscale}}%
    \fi%
  \else%
    \setlength{\unitlength}{\svgwidth}%
  \fi%
  \global\let\svgwidth\undefined%
  \global\let\svgscale\undefined%
  \makeatother%
  \begin{picture}(1,0.33229762)%
    \lineheight{1}%
    \setlength\tabcolsep{0pt}%
    \put(0.60700067,0.31190363){\color[rgb]{0,0,0}\makebox(0,0)[lt]{\lineheight{1.25}\smash{\begin{tabular}[t]{l}Gain Augmentation\end{tabular}}}}%
    \put(0.14337994,0.31188877){\color[rgb]{0,0,0}\makebox(0,0)[lt]{\lineheight{1.25}\smash{\begin{tabular}[t]{l}Original\end{tabular}}}}%
    \put(0.63874615,0.13245816){\color[rgb]{0,0,0}\makebox(0,0)[lt]{\lineheight{1.25}\smash{\begin{tabular}[t]{l}Freq-MixStyle\end{tabular}}}}%
    \put(0.12201746,0.13210336){\color[rgb]{0,0,0}\makebox(0,0)[lt]{\lineheight{1.25}\smash{\begin{tabular}[t]{l}SpecAugment\end{tabular}}}}%
    \put(0,0){\includegraphics[width=\unitlength,page=1]{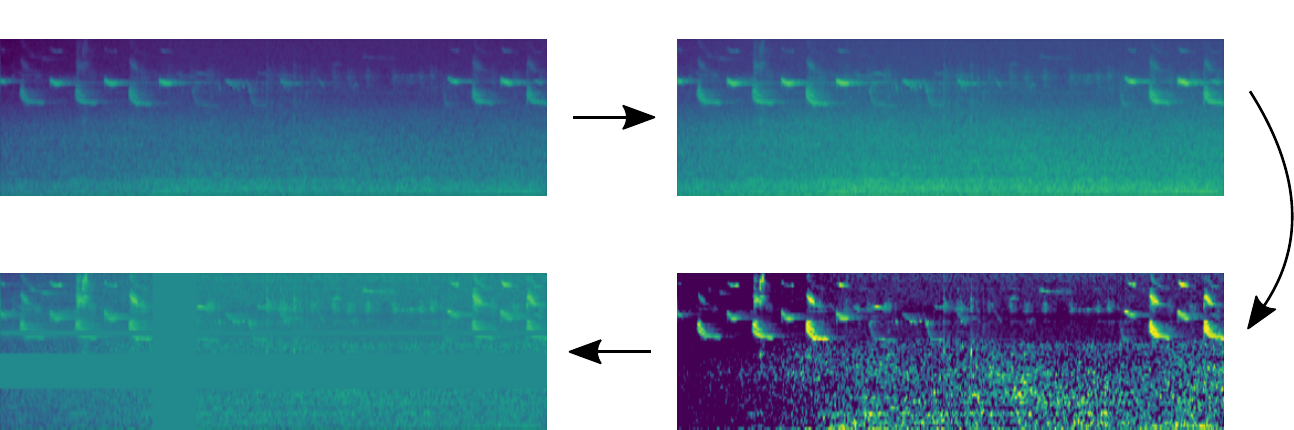}}%
  \end{picture}%
\endgroup%

}{%
  \caption{Overview of the audio augmentation pipeline.}%
  \label{fig:overview_audio_augment}
}
\capbtabbox{%
\begin{tabular}{@{}ll@{}}
\toprule
Augmentation     & Caption                                        \\ \midrule
Original         & \texttt{The rain pours down.}                           \\
Back Translation & \texttt{It rains cats and dogs.}                        \\
Insert           & \texttt{It {\color[HTML]{009901}tree} rains cats and dogs.}                        \\
Delete           & \texttt{It rains cats and {\color[HTML]{CB0000}\st{dogs}}.}                        \\
Swap             & \texttt{It {\color[HTML]{F56B00}and} cats {\color[HTML]{F56B00}rains} dogs.} \\
Synonym          & \texttt{It {\color[HTML]{F56B00}drizzles} cats and dogs.}                     \\ \bottomrule
\end{tabular}
}{%
  \caption{Overview of the text augmentation pipeline}%
  \label{tab:overview_text_augment}
}
\end{floatrow}
\end{figure*}

\section{Retrieval System}

Our model uses separate audio and caption embedding networks $\phi_a(\cdot)$ and $\phi_c(\cdot)$ to embed tuples of spectrograms and descriptions $\{(a_i, c_i)\}_{i=1}^{N}$ into a shared $D$-dimensional space in a manner that representations of matching audio-caption pairs are close. This behavior is achieved by contrastive training, which equalizes the embeddings of matching audio-caption pairs $(a_i, c_i)$, while pushing the representations of mismatching pairs $(a_i, c_{j; j \neq i})$ apart. The agreement between audio $a_i$ and description $c_j$ is estimated via the normalized dot product in the shared embedding space:
$$C_{ij} = \frac{\phi_{\textrm{a}}(a_{i})^T \cdot \phi_{\textrm{c}}(c_{j})}{\norm{\phi_{\textrm{a}}(a_{i})}^2 \norm{\phi_{\textrm{t}}(c_{j})}^2}$$
The similarity matrix $\mathbf{C} \in \mathbb{R}^{N \times N}$ holds the agreement of matching pairs on the diagonal and the agreement of mismatching pairs off-diagonal. We train the system using the NT-Xent \cite{NTxent} loss, which is defined as the average Cross Entropy ($\mathrm{CE}$) loss over the audio and text dimension; the ground truth is given by the identity matrix $\mathbf{I} \in \mathbb{R}^{N \times N}$:
$$\mathcal{L} = \frac{1}{2\cdot N} \sum_{i=1}^{N} \textrm{CE}(\mathbf{C}_{i*}, \mathbf{I}_{i*}) + \textrm{CE}(\mathbf{C}_{*i}, \mathbf{I}_{*i})$$

\label{sec:retrieval}

\section{Audio Augmentations}
To reduce overfitting of the audio embedding model and improve generalization, we employ three regularization techniques during training: Gain augmentation, MixStyle \cite{mixstyle, freq_mixstyle} along the frequency dimension (Freq-MixStype), and SpecAugment \cite{specaugment}. Figure \ref{fig:overview_audio_augment} gives an overview of the audio augmentation pipeline.\\

\textbf{Gain Augmentation} tries to make the model invariant changes in volume by randomly altering the loudness of the raw audio input signal. Volume manipulations are done by multiplying the waveform with factor $W$:
$$
w = 10^{(g/20)}
$$
The change in volume (in $\textrm{dB}$) is controlled with hyperparameter $g$; its value is randomly drawn from a uniform distribution in the range $[-g_\textrm{max}, g_\textrm{max}]$. \\

\textbf{SpecAugment} \cite{specaugment} randomly masks time and frequency stripes in the input spectrogram, thereby reducing the audio embedding model's reliance on specific input patterns. The number of stripes along the time and frequency dimensions is controlled via hyperparameters $n_f$ and $n_t$, respectively. Parameters $w_f$ and $w_t$ control the maximum width of the time and frequency stripes, respectively. The actual width and the offset of the stripes are chosen from a uniform distribution; masked values are replaced with zeros.  We omitted the warping transformation proposed in the original work as it is computationally expensive and reportedly only lead to marginal improvements.\\

\textbf{Freq-MixStyle} \cite{mixstyle} aims to transfer device-style characteristics between recordings by exchanging statistics along the frequency dimension of spectrograms. To this end, the original spectrogram is first normalized to zero mean unit variance along the frequency dimension and then un-normalized with adjusted mean and standard deviation statistics. The adjusted statistics are a convex combination of the original statistics $(\mu_i, \sigma-i)$ and the statistic of a randomly selected spectrogram $(\mu_j, \sigma_j)$: 
$$\mu_{\textrm{new}} = \lambda \mu_i + (1 - \lambda) \mu_j$$
$$\sigma_{\textrm{new}} = \lambda \sigma_i + (1 - \lambda) \sigma_j$$
The coefficient $\lambda$ is drawn from a symmetric beta distribution in a manner that the original statistics always receive a higher weight:
$$\lambda \sim \textrm{Beta}(\alpha, \alpha)$$
$\alpha$ controls the shape of the Beta distribution. Freq-MixStyle is applied to each input example with a probability of $p_{\textrm{MS}}$.

\section{Text Augmentations}
We apply Back Translation \cite{backtranslation} and Easy Data Augmentation \cite{EDA} (in that order) to reduce overfitting of the sentence embedding model. Examples of these augmentations are given in Table \ref{tab:overview_text_augment}.\\

\textbf{Back Translation} (BT) \cite{backtranslation} introduces variation into the input sentence without changing its semantics by translating the input sentence to a foreign language and back to the source language. We translate the training captions from English to German, French, or Spanish, and back to English using Google Translate.\\

\textbf{Easy Data Augmentation} (EDA) \cite{EDA} chooses one of four word-level manipulations and applies the selected operation to each word with a certain probability (indicated in parenthesis): insertion of a random word ($p_{\textrm{ins}}$), deletion ($p_{\textrm{del}}$), swap with another word in the sentence ($p_{\textrm{swp}}$), or replacement with a synonym according to WordNet \cite{wordnet} ($p_{\textrm{syn}}$). EDA is applied with a probability of $p_{\textrm{EDA}}$.

\section{Experiments}
We first established a baseline without augmentation and then conducted a series of experiments to investigate the impact of using pretrained weights for the audio embedding model, augmenting the audio and text inputs, and pretraining on AudioCaps. We further investigate the impact of all augmentation methods separately in an ablation study. The model architecture and the exact experimental setup are discussed below.

\subsection{Dataset \& Input Features}
We trained our proposed system on ClothoV2.1 \cite{clotho}, which contains $10$-$30$ second long audio recordings sampled at $32$kHz and five human-generated captions for each recording. We used the training, validation, and test split into 3839, 1045, and 1045 examples, respectively, as suggested by the dataset's creators. To make processing in batches easier, we zero-padded all audio snippets to the maximum audio length in the batch. The resulting waveforms were converted to $64$-bin log-MEL spectrograms using a $1024$-point FFT ($32$ms) and hop size of $320$ ($10$ms). The audio features were normalized via batch normalization \cite{batchnorm} along the frequency dimension before feeding them into the CNN10 embedding model. The input sentences were pre-processed by converting all characters to lowercase and removing punctuation. The resulting strings were tokenized with the WordPiece tokenizer \cite{wordpiece}, padded to the maximum sequence length in the batch, and truncated to 32 tokens.

\subsection{Audio Embedding Model}
We used a slightly modified version of the popular CNN10 architecture \cite{panns} to embed spectrograms into the 1024-dimensional audio-caption space. The architecture is detailed in Table \ref{tab:CNN10}. The network aggregates the output after the last convolutional block over the frequency and time dimensions and transforms the result with a two-layer neural network. The audio embedding model has approximately 9 Million parameters. We chose this simple architecture because it allowed us to train on a single customer-grade GPU with reasonable batch size. For the experiments with pretrained audio embedding model parameters, we transferred the weights of the convolutional blocks from a custom pretrained audio tagger and randomly initialized the fully-connected layers. The data set used for pretraining the embedding model, AudioSet \cite{audioset}, contains approximately 2 Million ten-second audio recordings labeled for 527 hierarchically organized classes. We used the pre-defined split of AudioSet into a large, unbalanced set for training and two smaller, more balanced sets for validation and testing. Pretraining of the embedding model was done as described in \cite{panns}.

\begin{table}[ht]
\centering
\begin{tabular}{@{}c@{}}
\toprule
CNN10               \\ \midrule
$2 \times (3 \times 3) @ 64$, BN, ReLU  \\
Pool $(2 \times 2)$                     \\
$2 \times (3 \times 3) @ 128$, BN, ReLU \\
Pool $(2 \times 2)$                     \\
$2 \times (3 \times 3) @ 256$, BN, ReLU \\
Pool $(2 \times 2)$                     \\
$2 \times (3 \times 3) @ 512$, BN, ReLU \\
Pool $(2 \times 2)$                     \\ \midrule
Frequency Pooling (mean)                     \\
Time Pooling (average of mean and max )                       \\\midrule

FC 2048, ReLU \\
FC 1024       \\ \bottomrule
\end{tabular}

\caption{The architecture of the audio embedding model (CNN10).}
\label{tab:CNN10}
\end{table}

\subsection{Text Embedding Model}

We used a pretrained BERT model \cite{bert} ('bert-base-uncased') to generate embeddings for the audio captions. BERT is a bi-directional self-attention-based sentence encoder that was pretrained on BookCorpus \cite{bookcorpus} and WikiText datasets \cite{wikitext} for masked language modeling and next sentence prediction. The learned semantic representations proved effective in multiple downstream tasks. We projected the output vector that corresponds to the class token into the shared audio-caption space by using a neural network with one hidden layer of size 2048 and ReLu activations. The text embedding model has approximately 112 Million parameters.

\subsection{Training \& Evaluation} \label{sec:train}
We train all variants of the proposed system on CLothoV2.1's training set, select hyperparameters according to the performance on the validation set, and report the final results on the test set in section \ref{sec:results}. Our main evaluation criterion was the mean Average Precision among the top-10 results (mAP) because this criterion takes the rank of the correct recording into account. We also report the recall among the top-1, top-5, and top-10 retrieved results. All results are averaged over three runs. Both embedding models were jointly optimized using gradient descent with a batch size of 30. We used the Adam update rule \cite{adam} for 50 epochs, set the initial learning rate to $10^{-4}$, and dropped it by a factor of $3$ every $10$ epochs. The hyperparameters of the optimizer were set to PyTorch's \cite{pytorch} defaults.

\subsection{Sequential Model-Based Optimization}
We performed sequential model-based optimization (SMBO) in the hyperparameter space of the audio and text augmentations to optimize the mAP-score on the validation set without manual tuning. Sequential Model-based Optimization (SMBO) utilizes the outcomes of prior experiments to build a surrogate model that estimates the relationship between validation-mAP and a given parameter configuration. Subsequent runs use this surrogate model to sample hyperparameter configurations from a distribution that is proportional to the expected mAP improvement. We initialized SMBO with ten runs using randomly chosen hyperparameters. After that, we performed 100 trials with hyperparameters sampled using the Tree-structured Parzen Estimator algorithm \cite{TPE}. To reduce the overall computation time, we stopped runs for which the mAP on the validation set did not increase for ten consecutive epochs. Table \ref{tab:smbo_hyperparam} defines the hyperparameter search space for the SMBO. 

\section{Results \& Discussion} 
\label{sec:results}
The results of our experiments are summarized in Table \ref{tab:results} and discussed in the following section.

\begin{table}[h]
\centering
\begin{tabular}{@{}lrrrr@{}}
\toprule
               & \multicolumn{1}{c}{R@1} & \multicolumn{1}{c}{R@5} & \multicolumn{1}{c}{R@10} & \multicolumn{1}{c}{mAP@10} \\ \midrule
DCASE baseline  &    $3.50$       & $11.50$         & $19.50$         & $7.50 \pm 0.00$   \\
baseline         &   $6.63$      & $20.06$         & $31.52$         & $12.53 \pm 0.08$  \\
+ AudioSet pretraining &    $13.18$      & $35.30$         & $48.61$         & $22.80 \pm 0.29$  \\ 
+ augmentations             &    $14.50$      & $37.24$         & $51.04$         & $24.27 \pm 0.19$  \\ 
+ AudioCaps pretraining            &    $14.34$      & $38.12$         & $52.04$         & $24.57 \pm 0.15$  \\ \bottomrule

\end{tabular}
\caption{Audio retrieval performance of the DCASE baseline and the custom system in four variants.}
\label{tab:results}
\end{table}

\subsection{Baseline}
The performance of our custom baseline system, which was trained with randomly initialized audio embedding model parameters and without augmentation, is given in Table \ref{tab:results}. The resulting system's mAP is 5 pp. higher than the mAP of the DCASE baseline system \cite{dcase2022_task6b} which we attribute to the more powerful text embedding model (we used BERT \cite{bert} instead of Word2Vec \cite{word2vec}).


\subsection{AudioSet Pretraining}

Next, we investigated the impact of using pretrained weights to initialize the audio embedding model. To that end, we retrained our baseline system but transferred the initial audio embedding model parameters from a CNN10 pretrained for tagging on AudioSet. The resulting model is approximately 10 pp. mAP better than the system that used randomly initialized weights for the audio encoder. This confirms that using pretrained audio embedding models is an effective strategy to alleviate the data scarcity problem.



\subsection{Augmentations}
We build upon the system that uses the AudioSet pretrained embedding model and perform SMBO to find a good hyperparameter configuration. The resulting best hyperparameters on the validation set and the performance on the test set are given in Tables \ref{tab:smbo_hyperparam} and \ref{tab:results}, respectively. The best configuration found suggests that the parameter which controls the frequency of EDA is superfluous as the best value is very close to one. Synonym replacement appears beneficial, and future experiments should search for the optimal value for this parameter in a larger range. The probability of swapping and inserting random words is close to zero, suggesting that these two transformations are less beneficial or even detrimental. All in all, we observed an absolute improvement of approximately 1.5 pp. mAP when training with text and audio augmentations.

\begin{table}[h]
\centering

\begin{tabular}{@{}lllrr@{}}
\toprule
                                                 & Augmentation                   & Parameter          & Range               & best     \\ \midrule
\multirow{6}{*}{\rotatebox[origin=c]{90}{Text}}  & \multirow{5}{*}{EDA}           & $p_{\textrm{EDA}}$ & $[0,1.0]$           &  $.9936$        \\
                                                 &                                & $p_\textrm{syn}$   & $[0,0.3]$           &  $.2962$        \\
                                                 &                                & $p_\textrm{swp}$   & $[0,0.3]$           &  $.0085$        \\
                                                 &                                & $p_\textrm{ins}$   & $[0,0.3]$           &  $.0269$        \\
                                                 &                                & $p_\textrm{del}$   & $[0,0.3]$           &  $.1944$        \\
                                                 & Backtranslation                & $p_\textrm{bt}$    & $[0,1.0]$           &  $.1812$        \\ \midrule
\multirow{7}{*}{\rotatebox[origin=c]{90}{Audio}} & \multirow{4}{*}{SpecAugment}   & $n_{\textrm{f}}$   & $\{0,1 \}$          &  $1$            \\
                                                 &                                & $w_{\textrm{f}}$   & $\{1, \dots, 32 \}$ &  $4$            \\
                                                 &                                & $n_{\textrm{t}}$   & $\{0, \dots, 8 \}$          &  $7$            \\
                                                 &                                & $w_{\textrm{t}}$   & $\{1, \dots, 64 \}$ &  $58$           \\
                                                 & Audio Gain                     & $g_\textrm{max}$                & $\{0, \dots, 6\}$   &  $3$            \\ 
                                                 & \multirow{2}{*}{Freq-MixStyle} & $p_\textrm{MS}$    & $[0,1.0]$           &  $.1045$        \\
                                                 &                                & $\alpha$           & $[0,1.0]$           &  $.8286$      \\ \bottomrule
\end{tabular}
\caption{Hyperparameter search space for the sequential model-based optimization, and the best configuration found.}
\label{tab:smbo_hyperparam}
\end{table}

\subsection{AudioCaps Pretraining}
We hypothesized that pretraining the retrieval system on additional audio-caption pairs could further improve audio-retrieval results; we, therefore, pretrained the system (with AudioSet pretraining and augmentations) on the $46K$ training examples in AudioCaps \cite{audiocaps}. To this end, we used the same training procedure as described in Section \ref{sec:train} for pretraining and fine-tuning but decreased the initial learning rate for fine-tuning by a factor of 10. Table \ref{tab:results} gives the results. Pretraining on AudioCaps resulted only in a marginal improvement of 0.3pp mAP, which suggests that using AudioCaps for transfer learning in this naive way has no significant impact.

\subsection{Ablation Study: Augmentations}
Based on the previous results, we performed another ablation study to investigate the effect of the audio and text augmentations. To that end, we re-trained the system with AudioSet pretraining and augmentations twice: once without audio augmentations and once without text augmentations. The results are given in Table \ref{tab:augmentations}. Using all augmentations gave the best results. We observed a drop of 1.1 and 0.5 pp mAP without text and audio augmentations, respectively. This might indicate that the text augmentations have a larger impact than the audio augmentations, which might be caused by the large difference in trainable parameters between the sentence and audio embedding models. \\

To isolate the effect of each individual augmentation method, we further re-trained the system (with AudioSet pretraining and augmentations) in five variants, always leaving out one of the augmentation methods. The results are summarized in Table \ref{tab:augmentations}. The text augmentations have the largest impact, which is in line with the previous results: Leaving out EDA and BT reduced the mAP by 1.0 and 0.7 pp., respectively. Eliminating SpecAugment and Freq-MixStyle reduced the performance by 0.7 and 0.6 pp., respectively. Gain augmentation seems to have the least impact: eliminating it reduced the mAP by only 0.2pp.

\begin{table}[h]
\centering
\begin{tabular}{@{}lrrrr@{}}
\toprule
               & \multicolumn{1}{c}{R@1} & \multicolumn{1}{c}{R@5} & \multicolumn{1}{c}{R@10} & \multicolumn{1}{c}{mAP@10} \\ \midrule
SMBO            &    $14.50$      & $37.24$         & $51.04$         & $24.27 \pm 0.19$  \\ \bottomrule
no audio aug    &    $13.88$      & $36.94$         & $51.06$         & $23.74 \pm 0.16$   \\
no text aug     &    $13.12$      & $35.77$         & $49.25$         & $22.91 \pm 0.08$  \\ \bottomrule
no SpeAugment       &    $13.50$      & $36.60$         & $50.91$         & $23.53 \pm 0.20$  \\
no FreqMixStyle    &    $13.61$      & $36.91$         & $50.69$         & $23.62 \pm 0.14$  \\ 
no Gain Augment &    $14.84$      & $37.81$         & $50.95$         & $24.05 \pm 0.26$  \\
no BT           &    $13.61$      & $36.38$         & $50.00$         & $23.43 \pm 0.18$   \\ 
no EDA          &    $13.33$      & $37.02$         & $49.94$         & $23.27 \pm 0.03$   \\\bottomrule
\end{tabular}
\caption{Results of the ablation study on data augmentation.}
\label{tab:augmentations}
\end{table}

\section{Conclusion}
This study set out to investigate transfer learning and data augmentation strategies to alleviate the data scarcity problem in natural-language-based audio retrieval. Our research has shown that using pretrained audio and text embedding models greatly increases the retrieval performance on ClothoV2. We enriched this already well-performing retrieval system with a range of augmentation methods and showed that augmenting both text and audio inputs significantly reduces overfitting. Finally, we further found that pretraining on AudioCaps only leads to non-significant improvements.
\section{ACKNOWLEDGMENT}
\label{sec:ack}
The LIT AI Lab is financed by the Federal State of Upper Austria.

\bibliographystyle{IEEEtran}
\bibliography{template}

\end{sloppy}
\end{document}